\documentclass[conference]{IEEEtran}
\usepackage{cite}
\usepackage{amsmath,amssymb,amsfonts}
\usepackage{amsthm}
\usepackage{algorithmic}
\usepackage{graphicx}
\usepackage{textcomp}
\usepackage{float}
\usepackage{xcolor}
\usepackage{multirow}
\usepackage{subfigure}
\usepackage{adjustbox}
\usepackage{stfloats}
\usepackage{balance}
\usepackage{algorithm2e}

\def\BibTeX{{\rm B\kern-.05em{\sc i\kern-.025em b}\kern-.08em T\kern-.1667em\lower.7ex\hbox{E}\kern-.125emX}}
\begin{document}
\SetKwComment{Comment}{/* }{ */}
\RestyleAlgo{ruled}
\title{\LARGE {Physical Layer Authentication for LEO Satellite Constellations}\\
{}

}

\author{
 \IEEEauthorblockN{ Ozan Alp Topal\IEEEauthorrefmark{1}\IEEEauthorrefmark{2}, 
and Gunes Karabulut Kurt\IEEEauthorrefmark{1}\IEEEauthorrefmark{3}}

\IEEEauthorblockA{ \IEEEauthorrefmark{1} \textit{Department of Electrical Engineering and Communications}, \textit{Istanbul Technical University}, Istanbul, 34469 Turkey }

\IEEEauthorblockA{ \IEEEauthorrefmark{2} \textit{School of Electrical Engineering and Computer Science}, \textit{KTH Royal Institute of Technology}, Stockholm, 100-44 Sweden}

\IEEEauthorblockA{\IEEEauthorrefmark{3} \textit{Department of Electrical Engineering} 
\textit{Polytechnique Montr\'eal, }
Montr\'eal, QC, Canada \\ E-mail: oatopal@kth.se,
gunes.kurt@polymtl.ca}
}

\maketitle

\begin{abstract}
Physical layer authentication (PLA) is the process of claiming identity of a node based on its physical layer characteristics such as channel fading or hardware imperfections. In this work, we propose a novel PLA method for the inter-satellite communication links (ISLs) of the LEO satellites. In the proposed PLA method, multiple receiving satellites validate the identity of the transmitter by comparing the Doppler frequency measurements with the reference mobility information of the legitimate transmitter and then fuse their decision considering the selected decision rule. Analytical expressions are obtained for the spoofing detection probability and false alarm probability of the fusion methods. Numerically obtained high authentication performance results pave the way to a novel and easily integrable authentication mechanism for the LEO satellite networks. 
\end{abstract}


\section{Introduction}
As the demand for connectivity boosts globally, space networks have become the next frontier in wireless communication. By providing continuous global and regional coverage, the space networks promise to support the connected \textit{everytime-everywhere} vision of 6G networks. Developing an integrated satellite and terrestrial network architecture is critical for boosting industries such as logistics, mining, agriculture, and defense. To obtain global connectivity vision, constellations which consist of thousands of satellites are needed to be established. Thanks to recent advances in rocket launch platforms, and the availability of the dedicated satellite spectrum, the number of satellites being launched into low Earth orbit (LEO) has increased 30 percent every year since 2012 \cite{Spectrum}. LEO satellites are roughly under 50 kilograms. Their limited size necessitates scalability in every operation that they initiate. Locations and velocities of the satellites are controlled by terrestrial units \cite{Spectrum}. Considering these characteristics, LEO satellite constellations act as space cyber-physical networks, where the control and connectivity of many low-cost and software-enabled controllable devices constitute the main priority \cite{IoT_space}.
  \begin{figure}[tbh]
  	\centering
 	\includegraphics[width=0.87\linewidth]{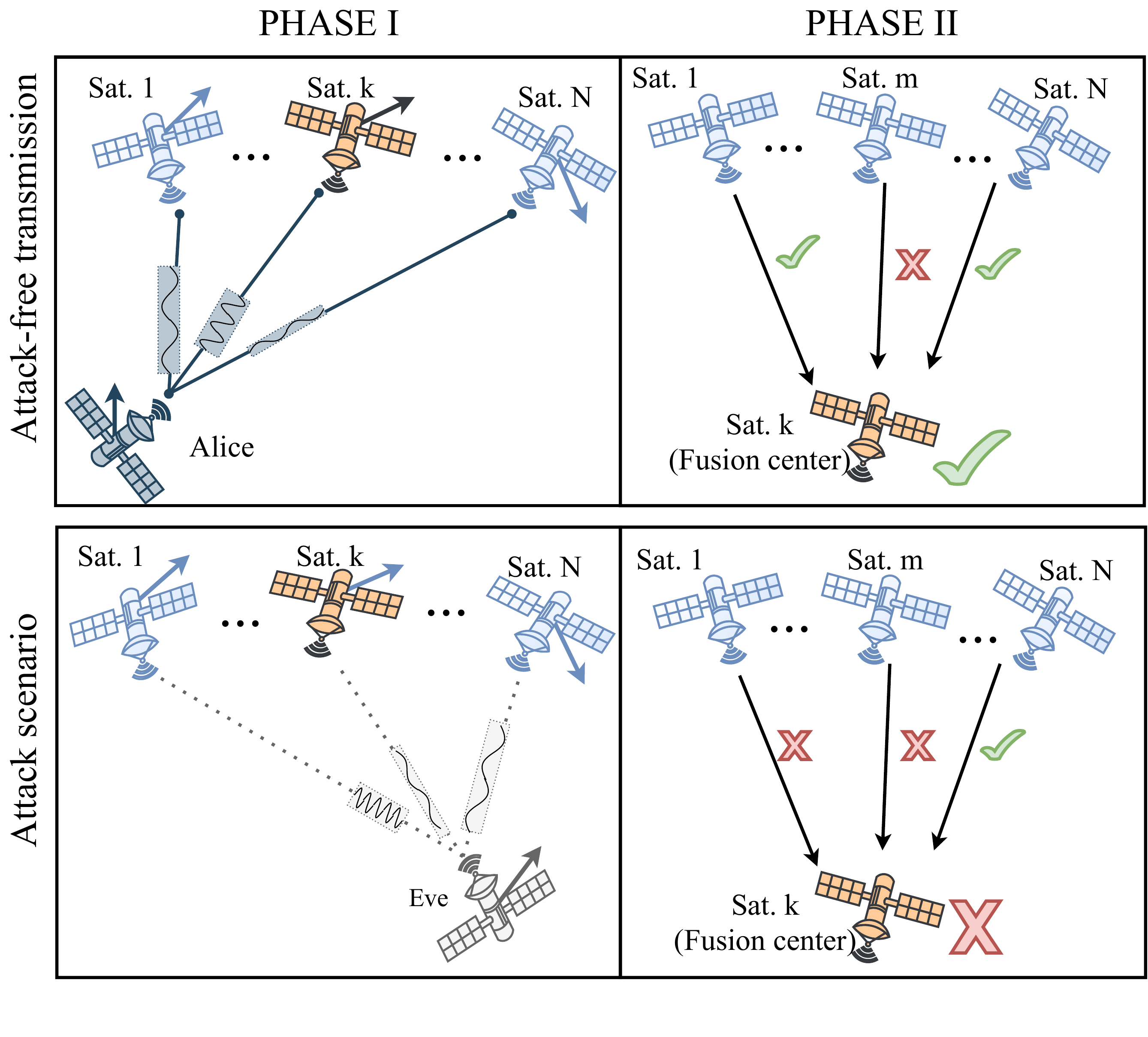}
 	\caption{Physical layer authentication scheme based on Doppler frequency shifts. At Phase I, each receiving satellite makes individual decisions. At phase II, these decisions are collected at fusion center to make the joint decision of the authentication. }
 	\label{fig:system}
 	\vspace{-0.5cm}
 \end{figure}

The open nature of the wireless communication channel constitutes security breaches for the space networks as in the wireless cyber-physical systems \cite{Topal2020}. One of the main concerns is spoofing attacks, where an active attacker tries to impersonate the legitimate transmitter to 	infiltrate and falsify the network. Therefore, any transmitter should prove its legitimacy to the receiver prior to starting the message transmission. Authentication is the act of proving this assertion, where it is conventionally obtained by cryptographic key-based methods at higher than the physical layer \cite{phy_auth}. The applicability of these methods would require a key management architecture for satellite or spacecraft networks, which is impractical considering the low-complexity requirement of the LEO satellite networks. Recently, the utilization of physical characteristics of the wireless channel such as channel fading or the characteristics of the transmitting device such as the non-linearity of the power amplifier has been proposed to provide an additional identity based on the physical layer. Named as physical layer authentication (PLA), these methods enable unclonable identity for wireless devices \cite{phy_auth}. Utilizing only PLA would create scalable authentication mechanisms for the low-power, low-chip area devices. Alternatively, utilizing PLA with upper layer authentication schemes would enable multi-factor authentication, and strengthens network security. 

Considering the dominant line of sight path in inter-satellite links (ISLs), the utilization of the channel fading based characteristics as a digital fingerprint becomes impractical. Although hardware imperfection based authentication might be utilized, device characteristics for such an implementation require  further investigation \cite{phy_auth}. {Characteristically}, ISLs  suffer from the high mobility of the LEO satellites, where this mobility reflects a dominant Doppler frequency shift \cite{Spectrum} in the received signal. Since the velocity and location information of every launched satellite is known from space and terrestrial networks, the LEO satellites can easily calculate the reference Doppler frequency shift values for any transmitting satellite. Consequently, LEO satellites can validate any transmitting user by comparing the measured Doppler shift from the received signal and the reference Doppler shift. Motivated by this idea, we propose a novel two-phase PLA scheme for the LEO satellites by utilizing Doppler frequency shifts as the digital fingerprints of the users. In the first phase, each satellite compares the measured Doppler frequency shifts with the reference values and individually decides on the identity of the transmitting user. In the second phase, they fuse their decisions and generate the final authentication decision. The contributions of this work can be listed as the following:
\begin{itemize}
\item We harness the Doppler frequency shift as the novel source of channel characteristics in the PLA literature to differentiate a spoofing attacker from the legitimate node. 
\item The proposed method does not require a secure channel to share the reference authentication source contrary to the state-of-the-art physical layer authentication methods. Since all satellites are obliged to share their mobility information for safety reasons, the Doppler shift resulted from this  mobility can easily be calculated and utilized as a reference.
\item By combining nominal power spectral density sample (NPSDS) decisions made by multiple satellites, in perfect estimation case with more than 6 satellites, the spoofer can be identified even if the attacker change the power and phase properties of its signal. For the imperfect NPSDS estimation case, we derive the spoofing detection probability and false alarm probability for individual decisions.
\item Through numerical analyses,  we compare the authentication performances of different decision fusion rules for the imperfect estimation case. The applicability of the proposed PLA scheme for the LEO satellite networks is shown through numerical results. 

\end{itemize}

 \subsection{Related Works}
 PLA methods can be classified into three categories device characteristics-based authentication, watermarking authentication, and channel based authentication. In device-characteristics-based authentication utilize hardware impairments as unique identifiers \cite{phy_auth}. In \cite{cfo}, the authors utilize carrier-frequency-offset (CFO) from hardware impairments for PLA.  
 Channel-based PLA methodologies utilize various fading channel characteristics such as channel state information (CSI) \cite{channel}, received signal strengths (RSSs) \cite{rss}, and angle of arrivals (AoA) \cite{aoa}. These systems are also extended to the multiple antenna transmitters and receivers \cite{mmimo}. As a novel source of identity, our proposed method utilizes the mobility of the transmitter as a digital fingerprint. From this perspective, our method can be classified under channel-based PLA methods with a unique channel characteristic. Due to the lack of multipath scattering for LEO satellites, fading based channel methods become impractical for the considered system. High mobility of the LEO satellites ensure continuous high Doppler frequency shift, and consequently a resource for the proposed PLA method. Furthermore, the proposed method does not require a feedback channel since all LEO satellites must share their mobility information with each other before service.
 
 \subsection{Notation}
 Scalar variables are denoted by italic symbols, vectors are
 denoted by boldface symbols. $\rho(X)$ denotes the probability density function (pdf) of the random variable $X$. $\mathbf{a}^T$ denotes the transpose of the vector $\mathbf{a}$. $||\mathbf{a}||$ denotes the Euclidean norm of the vector $\mathbf{a}$. $\gamma(x,y)$ denotes the lower-incomplete Gamma function, and $\Gamma(x,y)$ denotes the upper incomplete gamma function. $\Gamma(n)=(n-1)!$ denotes the Gamma function, where $(\cdot)!$ denotes the factorial operator. $\frac{\partial f}{\partial x}$ denotes the first-order partial derivative of $f$ with respect to $x$. $\mathbb{E}\{\cdot\}$ denotes the expectation operator. $A\cup B$ denotes the union of the sets $A$ and $B$. $||D||_1\geq \frac{N}{2}$, where $||D||_1$ denotes the $l_1$ norm of a vector $D$.

 \vspace{-0.2cm}
\section{System Model}
As illustrated in Fig. \ref{fig:system}, we consider a group consisted of $N$ number of LEO satellites in the orbit, and a LEO satellite which tries to authenticate with the group. This satellite might be a legitimate node or a spoofing attacker, Eve, who tries to mimic Alice to authenticate with the group. By utilizing the proposed PLA method, the group of satellites try to verify if the transmitting satellite is Alice or Eve. The proposed PLA method can be divided into two phases. In the first phase, each satellite decides the identity of the transmitter. Then, they send their decisions to the fusion center, which makes a final authentication decision in the second phase. In a communication slot, either Alice or Eve is assumed to have accessed the channel. In each communication slot, $N$ number of LEO satellites are assumed to receive the transmitted message either from Alice or from Eve. The transmitted message from Alice at a time instant $t$ is denoted by $m(t)$, where $t=\{1,2,\ldots,T\}$. Here, $T$ denotes the length of the communication  slot. In this work, as the worst case scenario, we assume that Eve knows the transmitted symbol sequence and would try to transmit the same sequence to the satellites. Note that, record and replay attacks enable the attacker to obtain $m(t)$, and they are exhaustively worked in the literature \cite{record_replay}. Considering this scenario, each satellite tries to decide between two hypothesis, where
\begin{equation}
\begin{cases}
& \mathcal{H}_0 \text{: Alice transmits,} \\
& \mathcal{H}_1 \text{: Eve transmits.} \\
\end{cases}
\end{equation}
After deciding on the hypothesis, their decisions are collected at the fusion center to jointly decide the authentication. In the following, we present under which conditions the Doppler frequency measurements may utilized as the digital fingerprints for the transmitters. 

\vspace{-0.2cm}
\subsection{Doppler Frequency Shift as a Fingerprint}
Doppler frequency shift depends on the locations and relative velocities due to the velocity and location of Alice is already available to the receiving satellites, the proposed system model does not require additional signal transmission from a secure communication channel.
 
 \vspace{0.25cm}
\noindent\textbf{Proposition 1:} When $N\geq6$ the number of Doppler frequency observations at receiving satellites are adequate to identify the transmitter. 
\begin{proof}
In the following, we use very similar steps to \cite{doppler_velocity}. One main difference is their formulations are given in $\mathbb{R}^2$, where we consider the locations and velocities in $\mathbb{R}^3$. Another difference is that their observer sensors are static, where our observer nodes have individual high velocities that cannot be neglected. Let us denote the position of the transmitting satellite in Cartesian $\mathbb{R}^3$ as a vector $\mathbf{p_t}=[p_1, p_2, p_3]^T$, and the velocity of the transmitting satellite in Cartesian $\mathbb{R}^3$ as a vector $\mathbf{v_t}=[v_1, v_2, v_3]^T$. The position of the $i^{\text{th}}$ receiving satellite is denoted by $\mathbf{p_{r_i}}=[p_{1,i}, p_{2,i}, p_{3,i}]^T$, and the velocity of $i^{\text{th}}$ receiving satellite is denoted by $\mathbf{v_{r_i}}=[v_{1,i}, v_{2,i}, v_{3,i}]^T$ for $i=\{1,2,\ldots,N\}$. $f_i$ denotes the nominal Doppler frequency, where $f_i=\frac{c\omega_i}{f_c}$, where $\omega_i$ is the Doppler frequency observed at the $i^{\text{th}}$ satellite, $c$ is the speed of light in m/s, and $f_c$ is the carrier frequency of the transmitted signal.  Then, the nominal Doppler frequency at the $i^{\text{th}}$ satellite is given by
\begin{equation}
\begin{aligned}
f_i&=\frac{\mathbf{v^T_t}(\mathbf{p_t}-\mathbf{p_{r_i}})}{||\mathbf{p_t}-\mathbf{p_{r_i}}||}+\frac{\mathbf{v_{r_i}^T}(\mathbf{p_{r_i}}-\mathbf{p_{t}})}{||\mathbf{p_{r_i}}-\mathbf{p_t}||}, \\ &=\frac{(\mathbf{v_t}-\mathbf{v_{r_i}})^T(\mathbf{p_t}-\mathbf{p_{r_i}})}{||\mathbf{p_t}-\mathbf{p_{r_i}}||}.
\end{aligned}
\end{equation}

 \begin{figure}[tb]
	\centering
	\includegraphics[width=0.6\linewidth]{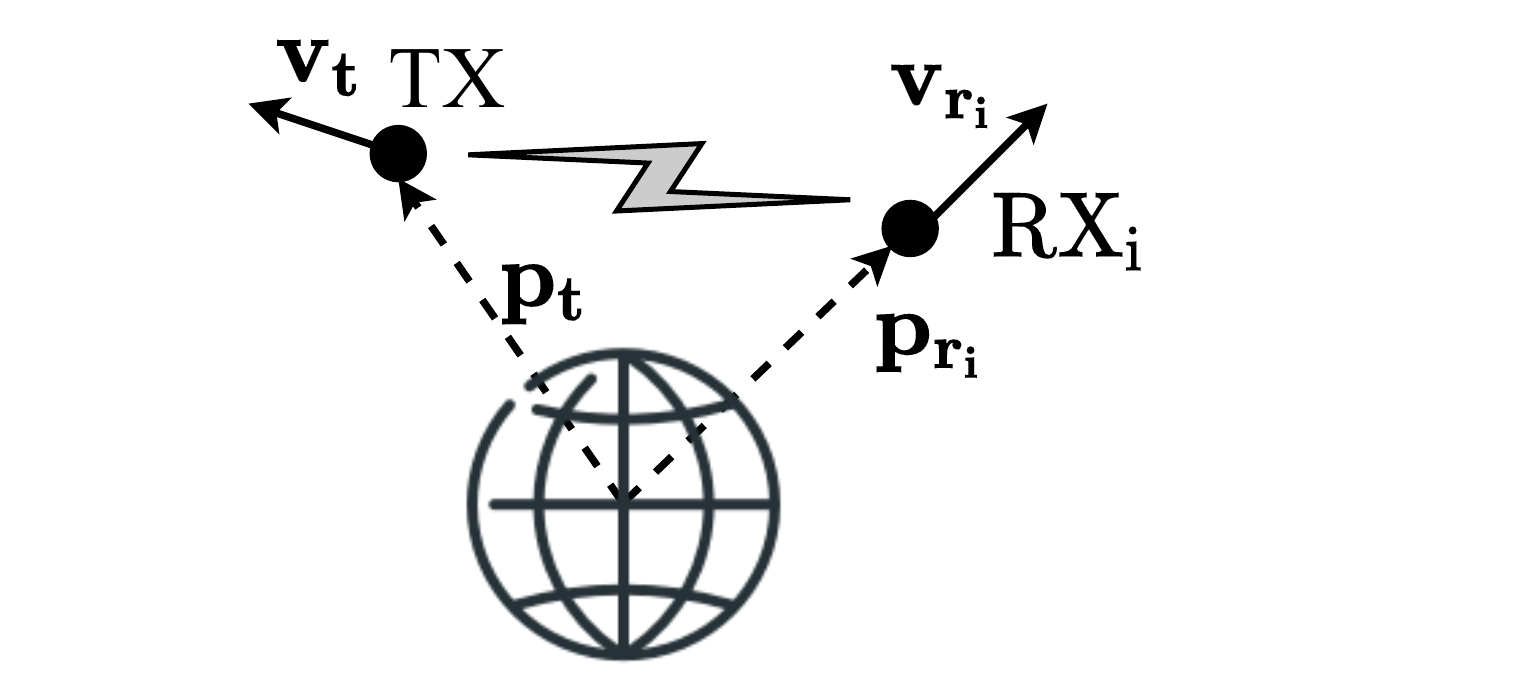}
	\caption{The visual definitions of the location and velocity vectors of the transmitter and the $i^{\text{th}}$ receiver. }
	\label{fig:doppler}
\end{figure}

Let concatenate this observations into a vector as $F(v_t,p_t)=[f_1,f_2,\ldots,f_N]^T$, and $\nabla F$ denotes the Jacobian of the nominal Doppler frequency observation vector, where 
\begin{equation}
\nabla F=  \begin{bmatrix}
\frac{\partial f_1}{\partial p_1} & \frac{\partial f_1}{\partial p_2}  & \frac{\partial f_1}{\partial p_3} & \frac{\partial f_1}{\partial v_1} & \frac{\partial f_1}{\partial v_2} & \frac{\partial f_1}{\partial v_3} \\ 
\vdots  & \vdots & \vdots & \vdots & \vdots & \vdots\\ 
\frac{\partial f_6}{\partial p_1} & \frac{\partial f_6}{\partial p_2}  & \frac{\partial f_6}{\partial p_3} & \frac{\partial f_6}{\partial v_1} & \frac{\partial f_6}{\partial v_2} & \frac{\partial f_6}{\partial v_3} 
\end{bmatrix}.
\end{equation}
The elements of this can be calculated by 
\begin{equation}
\footnotesize
\begin{aligned}
\frac{\partial f_i}{\partial p_1}&= \frac{\left(v_2(p_2-p_{i,2})+v_3(p_3-p_{i,3}) \right)(p_{i,1}-p_1)+v_1\left(p_{-1}\right)}{||\mathbf{p_t}-\mathbf{p_{r_i}}||^3}, \\
\frac{\partial f_i}{\partial p_2}&= \frac{\left(v_1(p_1-p_{i,1})+v_3(p_3-p_{i,3}) \right)(p_{i,2}-p_2)+v_2\left(p_{-2}\right)}{||\mathbf{p_t}-\mathbf{p_{r_i}}||^3}, \\
\frac{\partial f_i}{\partial p_3}&= \frac{\left(v_1(p_1-p_{i,1})+v_2(p_2-p_{i,2}) \right)(p_{i,3}-p_3)+v_3\left(p_{-3}\right)}{||\mathbf{p_t}-\mathbf{p_{r_i}}||^3}, \\
\frac{\partial f_i}{\partial v_1}&= \frac{p_1-p_{i,1}}{||\mathbf{p_t}-\mathbf{p_{r_i}}||}, 
\frac{\partial f_i}{\partial v_2}= \frac{p_2-p_{i,2}}{||\mathbf{p_t}-\mathbf{p_{r_i}}||}, 
\frac{\partial f_i}{\partial v_3}= \frac{p_3-p_{i,3}}{||\mathbf{p_t}-\mathbf{p_{r_i}}||},
\end{aligned}
\end{equation}
where $p_{-c}= \sum\limits_{C\neq c} (p_C-p_{i,C})^2$, $C=\{1,2,3\}$. By putting some generic $p_t$ and $v_t$ values, the non-singularity of the $\nabla F$ can be observed. The non-singularity of the Jacobian implies that for generic values for observations we do not have a continuous set of solutions. As proven in Proposition 2 of \cite{doppler_velocity}, the singularity of the Jacobian can be occured only when $(a)$ any observing satellite and the transmitter are collinear or $(b)$ $\mathbf{v_t}=0$. The condition $(a)$ cannot be satisfied considering the LEO satellite constellations, where any satellite cannot be collinear with all observing satellites. The condition $(b)$ also cannot be satisfied, since the satellites orbits continuously \cite{ISL}. Therefore, for our setup, we can guarantee that the Jacobian is singular, and we have 6 unknowns and 6 equations. Consequently, the solution of (1) is unique. 
\end{proof}
As a result of Proposition 1, two satellites with distinct velocities and positions cannot provide identical Doppler frequency observations at more than 6 observer satellites.

\subsection{Channel Model}
Radio-frequency (RF) communication links are assumed as available for the ISLs. Each LEO satellite is assumed to be equipped with a single omni-directional antenna. Considering very high speeds and dominant line-of-sight (LoS) component in the ISLs, the transmitted signals from Alice and Eve is assumed to be mainly affected by path loss fading, Doppler frequency shift, and the communication channel is modeled as an additive white Gaussian (AWGN) channel. The locations and velocities of Alice, and the receiving satellites are assumed to be available at all nodes, since the legitimate satellites need to follow the regulations. In this case the received signal at the $i^{\text{th}}$ satellite can be given by
\begin{equation}
y_i(t)= \begin{cases}
& h_{ai}(t)m(t)+w_i(t), \mathcal{H}_0, \\
& h_{si}(t)m(t)+w_i(t), \mathcal{H}_1,
\end{cases}
\label{eq:received_signal}
\end{equation}
 where $i=\{1,2,\ldots,N\}$. $h_{ki}(t)=\sqrt{l_{ki}}e^{j(\omega_{ki}\cos(\delta_i)t+\Phi_i)}$ denotes the channel fading from $k^\text{th}$ transmitter to the $i^{\text{th}}$ receiver satellite, where $k\in\{{a,s}\}$ respectively for Alice and Eve. $\omega_{ki}$ denotes the Doppler shift. $\delta_i$ is the angle of arrival of LoS component and $\Phi_i$ is the phase of the LoS component. $l_{ki}=E[|h_{ki}(t)|^2]$ denotes the channel power. We assume $m(t)\sim \mathcal{CN}(0,\sigma_m^2)$, where receivers only know the variance of the message signal. This assumption brings two main advantages. First, the receivers do not require a pilot message signal. Only knowing the power spectral density of the message signal is efficient for the proposed methodology. Second, the selected carrier frequency does not affect the estimation performance \cite{CRB_gaussian}.

\section{Physical Layer Authentication Model}
\begin{figure}[tb]
	\centering
	\includegraphics[width=0.7\linewidth]{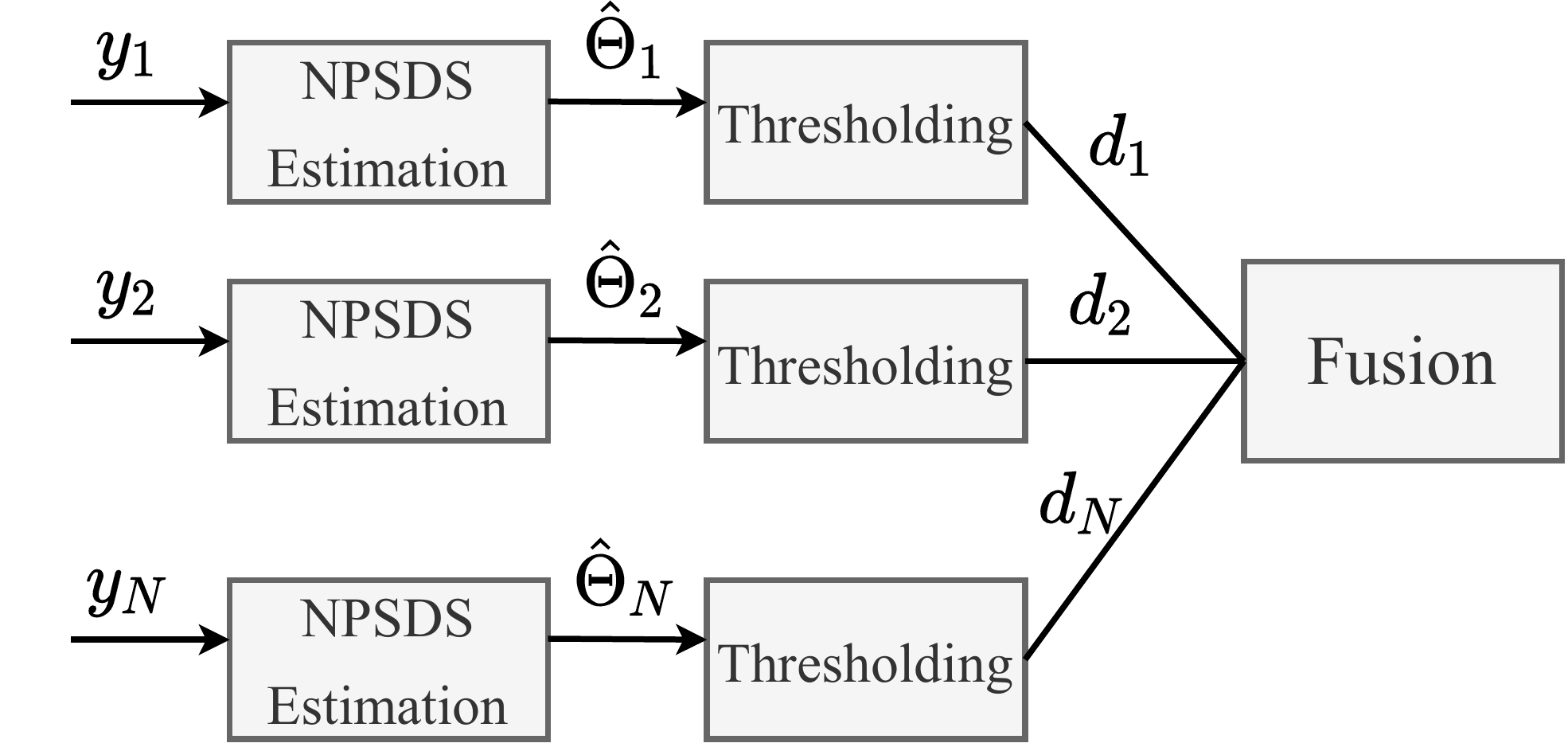}
	\caption{The block diagram of the proposed authentication scheme. }
	\label{fig:block_diagram}
\end{figure}

As illustrated in Fig. \ref{fig:block_diagram}, each satellite starts with estimating nominal power spectral density samples (NSPDS) to obtain the Doppler frequency information. Then, they utilize binary decision thresholding to decide whether Alice or Eve transmits. Finally, their decisions are collected at the fusion center to make a final authentication decision. In the following, we detail the each processes of the given block diagram. 
   
\subsection{NPSDS Estimation}

In NPSDS estimation, we follow similar manner to our previous works \cite{topal_space_1, topal_space_2}. Note that, we apply this operation for $N$ observing satellites.
Let us denote $x_{ki}(t)=h_{ki}(t)m(t)$, and $y_{ki}(t)=x_{ki}(t)+w_i(t)$. Note that, the information contained in $\mathbf{y_{ki}}$ is fully present in its discrete Fourier transform, 
$
Y_{ki}(t)=X_{ki}(t)+W_{ki}(t),
$
where $\mathbf{Y_{ki}}=\mathcal{F}\{y_{ki}\}=\left[Y_{ki}(1), Y_{ki}(2), \ldots, Y_{ki}(T) \right]$, $\mathbf{X_{ki}}=\mathcal{F}\{x_{ki}\}=\left[X_{ki}(1), X_{ki}(2), \ldots, X_{ki}(T) \right]$ and $\mathbf{W_{ki}}=\mathcal{F}\{w_{ki}\}=\left[W_{ki}(1), W_{ki}(2), \ldots, W_{ki}(T) \right]$. Since Gaussian processes are invariant against Fourier transform, the signal spectrum $\mathbf{X_{ki}}$, and the noise spectrum $\mathbf{W_{ki}}$, are also complex Gaussian, zero-mean, and orthogonal processes. The spectral samples ${Y_{ki}(t)}$ are mutually uncorrelated because of the assumed stationarity of $X_{ki}(t)$.

As stated in \cite{crbsystem}, the phase of $\mathbf{Y_{ki}}$ carries no information about the Doppler frequency, since $\mathbf{Y_{ki}}$ has been modeled as a stochastic process with the aforementioned properties. Hence, it is sufficient to consider the power spectrum of the received data as 

\begin{equation}
\begin{aligned}
\mathbf{S_{ki}} &= \left[S_{ki}(1), S_{ki}(2), \ldots, S_{ki}(T)\right] \\ 
&= \left[|Y_{ki}(1)|^2, |Y_{ki}(2)|^2, \ldots, |Y_{ki}(T)|^2\right].
\end{aligned}
\end{equation}

Since $Y_{ki}(t)$ is a complex Gaussian process, the probability density function of each sample ${S_{ki}}(t)$ under the condition of a particular Doppler frequency $\omega_{ki}$ is given by the exponential distribution \cite{crbsystem}:

\begin{equation}
\rho(S_{ki}(t);\omega_{ki})= \frac{1}{\Theta_{ki}(t)}\text{exp}\left(-\frac{S_{ki}(t)}{\Theta_{ki}(t)}\right), 
\end{equation}
where $\Theta_{ki}(t)$ denotes the NPSDS, and can be obtained by  
\begin{equation}
\begin{aligned}
\Theta_{ki}(t)= \mathbb{E}\{{S_{ki}}(t)\} &= \mathbb{E}\{|X_{ki}(t)+W_{ki}(t)|^2\} \\
&=\mathbb{E}\{|X_{ki}(t)|^2\}+\mathbb{E}\{|W_{ki}(t)|^2\}
\end{aligned},
\end{equation}
where $A^x_{ki}(t\Delta f-\omega_{ki})=\mathbb{E}\{|X_{ki}(t)|^2\}$, $A^n_{ki}=\mathbb{E}\{|K_{ki}(t)|^2\}$. Note that $A^x_{ki}(f)$ is the a priori known nominal power spectral density of the signal; $\Delta f$ is the frequency sampling interval; and $\omega_{ki}$ is the Doppler frequency shift. Considering $A^x_{ki}(f)$ is periodic with period $\Delta f$, and $A^n_{ki}$ is a constant, we can deduce that $\Theta_{ki}(t)$ is also periodic with $\Delta f$, and consequently we can drop $t$ and denote the NPSDS as $\Theta_{ki}=\Theta_{ki}(t), \forall t .$

Let us denote the estimated NPSDS at receiving node $i$ as $\hat{\Theta}_{ki}$. The maximum likelihood (ML) estimation of the parameter ${\Theta}_{ki}$ can be given as

\begin{equation}
\hat{\Theta}_{ki}= M^{ki}_s= {\frac{1}{T}\sum\limits_{t=1}^{T}S_{ki}(t)},
\end{equation}
where $ M^{ki}_s$ is the sample mean of the observed power spectral density samples \cite{topal_space_1}. By comparing the estimated NPSDSs with the reference NPSDS values, satellites decide whether or not the transmitted signal is coming from Alice. 

\subsection{Binary Hypothesis Test}
Considering the two hyphothesis given in (\ref{eq:received_signal}), estimated NPSDS at $i^{\text{th}}$ receiving node can be expressed by 

\begin{equation}
\hat{\Theta}_{i}= \begin{cases}
&\hat{\Theta}_{ai} \text{ , } \mathcal{H}_0, \\ 
&\hat{\Theta}_{si} \text{ , } \mathcal{H}_1,
\end{cases}
\end{equation}
where the pdf of $\hat{\Theta}_{ki}$ follows the Gamma distribution as in:

\begin{equation}
\rho(\hat{\Theta}_{ki})= \frac{1}{\Gamma(T)\Theta^T_{ki}T^{-T}}\hat{\Theta}^{T-1}_{ki}\exp(-\frac{\hat{\Theta}_{ki}T}{\Theta_{ki}}). 
\label{eq:pdf}
\end{equation}
In the following analysis, we will denote $\Theta_{si}=\beta_i\Theta_{ai}$, where $\beta_i$ is a positive finite real number.

\textit{Remark 1:} Let us assume that spoofing attacker emits a phase shifted version of the original message as $x_s(i)=m(t)e^{j\psi}$, the observed NPSDS values at receiving satellites would be $\Theta_{si}+\Theta_{\psi}$,  where it cannot be equal to $\Theta_{ai}$ unless all elements of $\Theta_{ai}$ is not equal. In other words, by changing transmitting signal power or phase, the attacker cannot spoof all observing satellites as long as these satellites are not identical. Therefore we utilize $\beta(i)$ as the deviation affect resulted from mobility of spoofer and any form of its attack.

Since each receiver knows the velocity and the location of Alice, they can calculate $\Theta_{ai}$ at any time instant. Therefore, each receiver can compare the estimated $\hat{\Theta}_i$ with the reference $\Theta_{ai}$, and make a decision on whether the channel is accessed by Alice or Eve based on this comparison. Let us $d_i$ denotes the decision of the $i^{\text{th}}$ satellite for a single communication slot, where $d_i=0$ and $d_i=1$ respectively for Alice and Eve has accessed the channel. The decision process at the $i^{\text{th}}$ satellite can be given as 
\begin{equation}
 \begin{cases}
 &|\hat{\Theta}_{i}-\Theta_{ai}|\leq \lambda_i \text{ , } d_i=0, \\ 
 &|\hat{\Theta}_{i}-\Theta_{ai}|> \lambda_i \text{ , } d_i=1,
 \end{cases}
\end{equation}
where $\lambda_i$ denotes the decision threshold for the $i^{\text{th}}$ satellite. 
In this case, the probability of spoofing detection at $i^{\text{th}}$ satellite can be expressed by $$\begin{aligned}
P_{d_i}&= Pr[|\hat{\Theta}_{i}-\Theta_{ai}|> \lambda_i|\mathcal{H}_1] \\ &=Pr[|\hat{\Theta}_{si}-\Theta_{ai}|> \lambda_i].
\end{aligned}$$
Considering the pdf expression given in (\ref{eq:pdf}) , $P_{d_i}$ becomes
\begin{equation}
\begin{aligned}
P_{d_i} &= \int_{\Theta_{ai}+\lambda_i}^{\infty}\rho(\hat{\Theta}_{si}) d\hat{\Theta}_{si} + \int_{-\infty}^{ \Theta_{ai}-\lambda_i}\rho(\hat{\Theta}_{si}) d\hat{\Theta}_{si}, \\
&= \frac{\gamma\left(T,\frac{T(\Theta_{ai}-\lambda_i)}{\Theta_{si}}\right)+\Gamma\left(T,\frac{T(\Theta_{ai}+\lambda_i)}{\Theta_{si}}\right)}{\Gamma(T)}.
\end{aligned}
\end{equation}

Similarly the probability of false alarm at the $i^{\text{th}}$ satellite can be expressed by
$P_{f_i}= Pr[|\hat{\Theta}_{i}-\Theta_{ai}|> \gamma_i|\mathcal{H}_0] =Pr[|\hat{\Theta}_{ai}-\Theta_{ai}|> \gamma_i].$
The closed form expression of the false alarm probability can be given by
\begin{equation}
\begin{aligned}
P_{f_i} &= \int_{\Theta_{ai}+\lambda_i}^{\infty}\rho(\hat{\Theta}_{ai}) d\hat{\Theta}_{ai} + \int_{-\infty}^{ \Theta_{ai}-\lambda_i}\rho(\hat{\Theta}_{ai}) d\hat{\Theta}_{ai}, \\
&= \frac{\gamma\left(T,\frac{T(\Theta_{ai}-\lambda_i)}{\Theta_{ai}}\right)+\Gamma\left(T,\frac{T(\Theta_{ai}+\lambda_i)}{\Theta_{ai}}\right)}{\Gamma(T)}.
\end{aligned}
\end{equation}

In the following analysis, to simplify the given expressions, we represent $\lambda_i=\alpha_i\Theta_{ai}$. Due to the semi-positive definition of the Gamma distribution, the feasible region for $\alpha_i$ becomes $0<\alpha_i<1$. Considering $\Theta_{si}=\beta_i\Theta_{ai}$ and $\lambda_i=\alpha_i\Theta_{ai}$, the spoofing detection probability and false alarm probability becomes 
\begin{equation}
\begin{aligned}
& P_{d_i}(\alpha_i,\beta_i,T)= \frac{\gamma\left(T,\frac{T(1-\alpha_i)}{\beta_{i}}\right)+\Gamma\left(T,\frac{T(1+\alpha_i)}{\beta_i}\right)}{\Gamma(T)}, \\
& P_{f_i}(\alpha_i,T)= \frac{\gamma\left(T,{T(1-\alpha_i)}\right)+\Gamma\left(T,{T(1+\alpha_i)}\right)}{\Gamma(T)}. 
\end{aligned}
\end{equation}
With given probabilities, each satellite determines optimum decision threshold by solving the following problem 
\begin{equation}
\begin{aligned}
\text{\textbf{P1:   }}\min_{\alpha_i}&\{ P_{f_i}+(1-P_{d_i})\}, \\
\text{s.t. } & 0<\alpha_i<1, T, \beta_{i}\in\mathbb{R^+}. \\
\end{aligned}
\end{equation} 
The analytical solution of this problem becomes intractable considering the complexity of the lower incomplete and upper incomplete Gamma functions. Therefore in Section IV, we numerically solve the problem to find the optimum threshold for each receiving satellite. As a result, obtained optimum threshold coefficient, $\alpha^*$, is inserted into the detection probability and false alarm probability expressions as $P^*_{d_i}=P_{d_i}(\alpha^*)$, $P^*_{f_i}=P_{f_i}(\alpha^*)$.  In the following, we address possible fusion mechanisms to jointly decide whether there is a spoofing attack or not. 

\subsection{Decision Fusion}
We consider three common methods in decision fusion. In Section IV, we compare the performance of these decision mechanisms in detail. Let us denote the vector composed of concatenated decisions gathered by N satellites as $D=[d_1,d_2,\ldots,d_N]$. $\mathbb{D}^+$ denotes the set of possible $D$ vector combinations, when the joint decision is that the spoofing attack is detected. Conversely, $\mathbb{D}^-$ denotes the set of possible $D$ vector combinations, when the joint decision is that Alice is detected, and $\mathbb{D}=\{\mathbb{D}^+ \cup \mathbb{D}^-\}$. The spoofing detection and false alarm probabilities for joint decision are respectively denoted by $\mathbb{P}_D$ and $\mathbb{P}_F$.
\subsubsection{OR Rule}
The first considered decision method is OR rule, where
$
\mathbb{D}^+=\mathbb{D}- [d_0=0, d_1=0, \ldots, d_N=0].
$
For OR rule, the spoofing detection probability and false alarm probability respectively become
\vspace{-0.05cm}
\begin{equation}
\mathbb{P}_D= 1-\prod_{i=1}^{N}(1-P^*_{d_i}), \text{ }
\mathbb{P}_F= 1-\prod_{i=1}^{N}(1-P^*_{f_i}).
\end{equation}
\vspace{-0.05cm}
As we detail in the following section, both detection and false alarm probabilities would be highest in OR rule compared with other methods.
\subsubsection{AND Rule}
The second considered decision method is AND rule. In this method, the spoofing decision is only positive for 
$
\mathbb{D}^+=[d_0=1, d_1=1, \ldots, d_N=1].
$
For AND rule, the spoofing detection probability and false alarm probability respectively become
\vspace{-0.05cm}
\begin{equation}
\mathbb{P}_D= \prod_{i=1}^{N}(1-P^*_{d_i}), \text{ }
\mathbb{P}_F= \prod_{i=1}^{N}(1-P^*_{f_i}).
\end{equation} 
\vspace{-0.05cm}
AND rule provides the least spoofing detection probability and the false alarm probability. 
\subsubsection{Majority Rule}
To compansate the spoofing detection and false alarm probabilities, finally we consider majority rule in decision fusion. In the majority rule at least $l= \left \lfloor  \frac{N}{2} \right \rfloor$ decisions should be positive to decide a spoofing attack existence. In this case the the spoofing detection probability and false alarm probability respectively become
\begin{equation}
\begin{aligned}
\mathbb{P}_D&= \sum_{d\in\mathbb{D}^+}{\left[\prod_{i=1}^{N}(P^*_{d_i})^{d_i}(1-P^*_{d_i})^{(1-d_i)}\right]}, \\
\mathbb{P}_F&= \sum_{d\in\mathbb{D}^+}{\left[\prod_{i=1}^{N}(P^*_{f_i})^{d_i}(1-P^*_{f_i})^{(1-d_i)}\right]}.
\end{aligned}
\end{equation}
Majority rule provides acceptable detection probability and low false alarm probability as discussed in the following section. The proposed PLA model is summarized in the Algorithm \ref{alg:1}.

\begin{algorithm}
	\caption{The proposed PLA approach.}\label{alg:1}
	\KwData{$\Theta_{a,1},\Theta_{a,2}, \ldots, \Theta_{a,N}, T, \mathbf{y}_1, \mathbf{y}_2, \ldots, \mathbf{y}_N $}
	\KwResult{$d_f$}
	$\mathbf{S}_{i} \gets |\mathcal{F}\{y_i\}|^2$\;
    $\hat{\Theta}_{i}\gets \frac{1}{T}\sum_{t=1}^{T} {S}_{i}(t)$  \;
    Solve P1 as in (17) $\forall i$ \; 
    Send $D=\left[d_1, d_2, \ldots, d_N\right]$ to the fusion center\;
	\eIf{$D\in \mathbb{D}^+$}{
	$d_f\gets 1 $\;
	}{$d_f\gets 0 $}
\end{algorithm}

\begin{figure}[t]
	\begin{center}	
		\subfigure[]{
			\label{fig:ROC1}
			\includegraphics[width=0.46\linewidth]{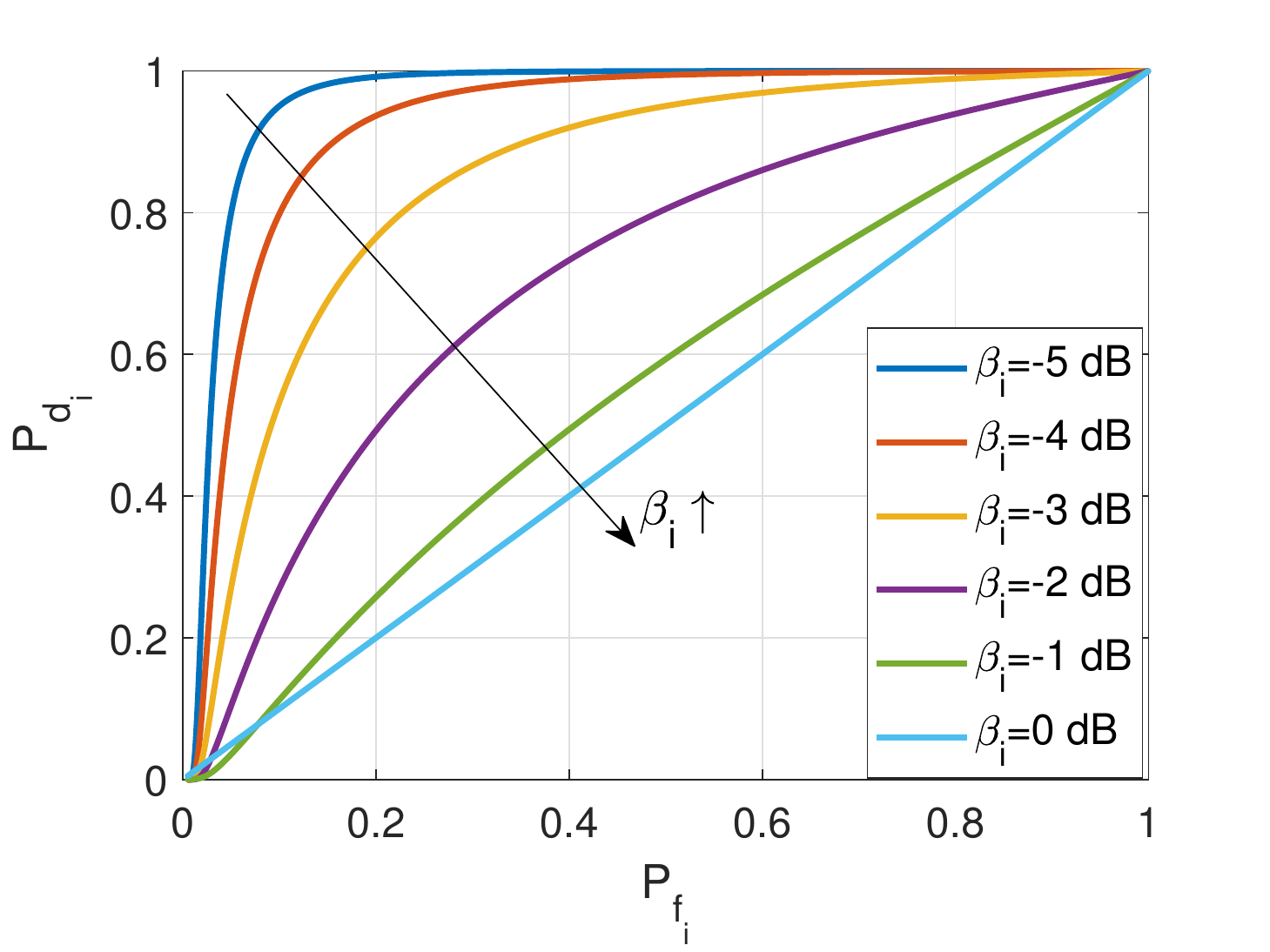} } 
		\subfigure[]{		
			\label{fig:ROC2}
			\includegraphics[width=0.46\linewidth]{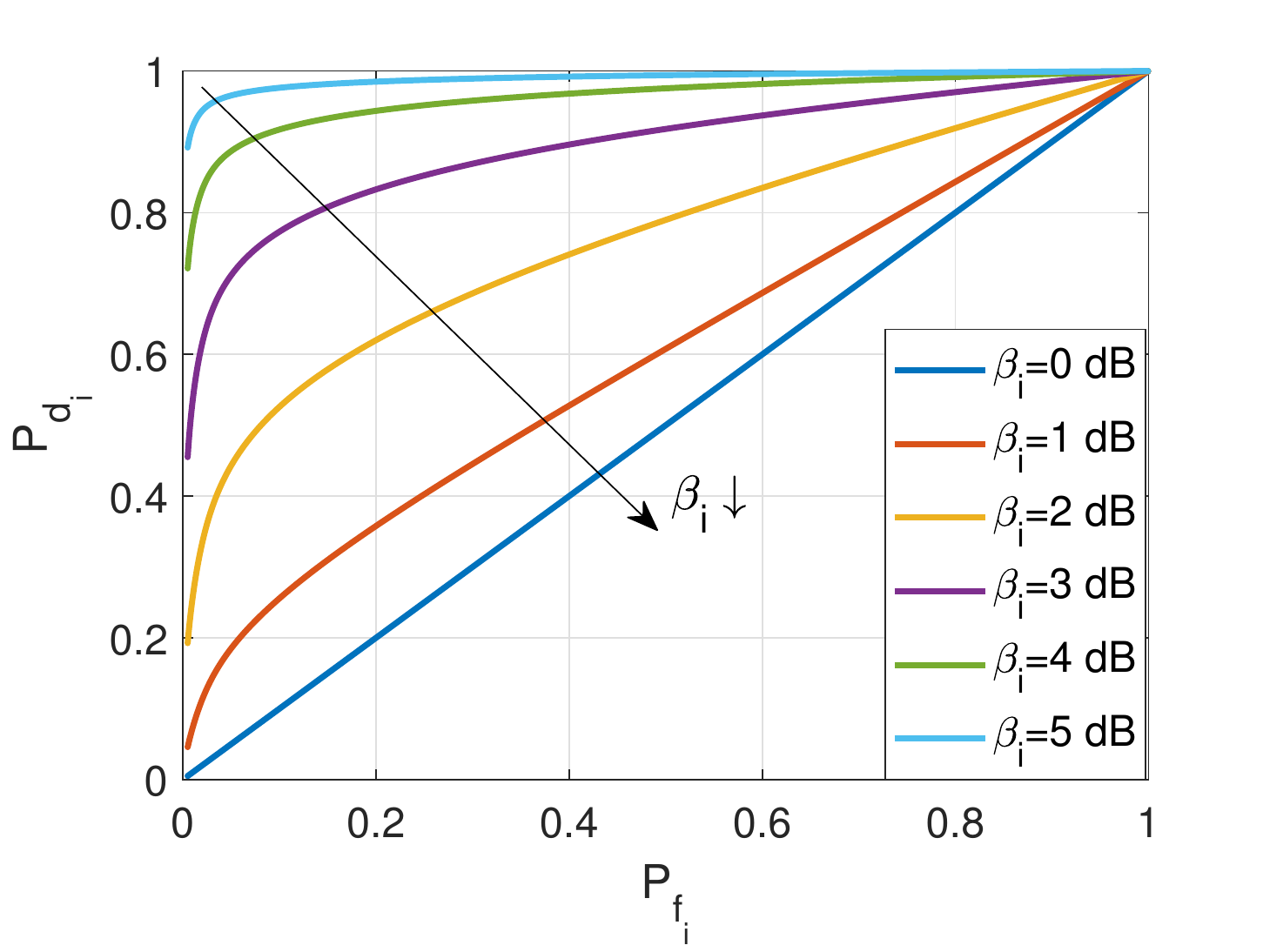} }
	\end{center}
		\vspace{-0.5cm}
	\caption{ROC curves at the $i^{\text{th}}$ receiving satellite for $0<\alpha<1$, and \newline (a) $0<\beta_i\leq1$, (b) $1<\beta_i$.}
	\label{fig:ROCs}
	\vspace{-0.5cm}
\end{figure}

\section{Numerical Results}
\begin{figure*}[tbh]
	\begin{center}	
		\subfigure[]{
			\label{fig:fusion_1}
			\includegraphics[width=0.3\textwidth]{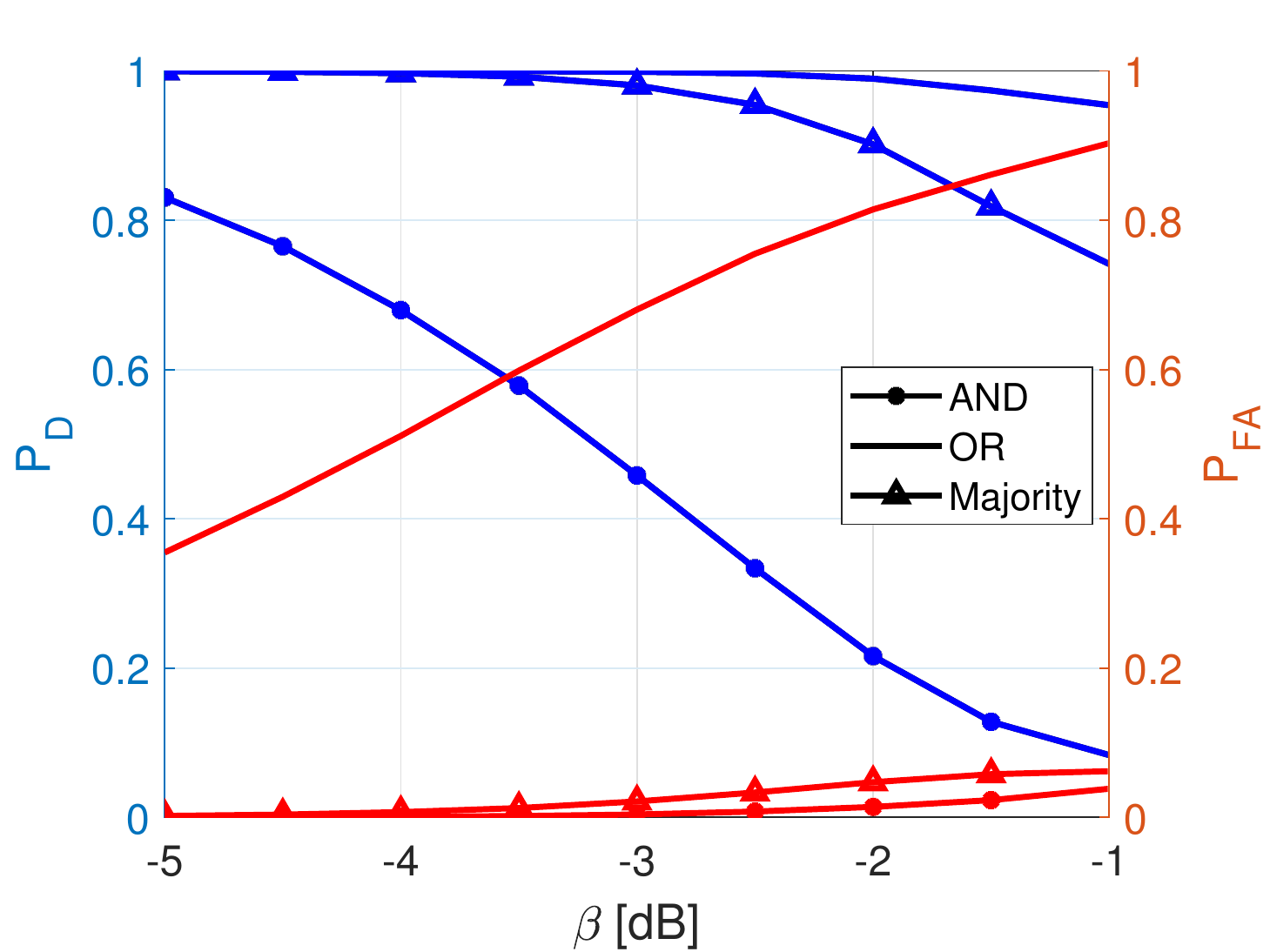} }
		\hfill
		\subfigure[]{
			\label{fig:fusion_2}
			\includegraphics[width=0.3\textwidth]{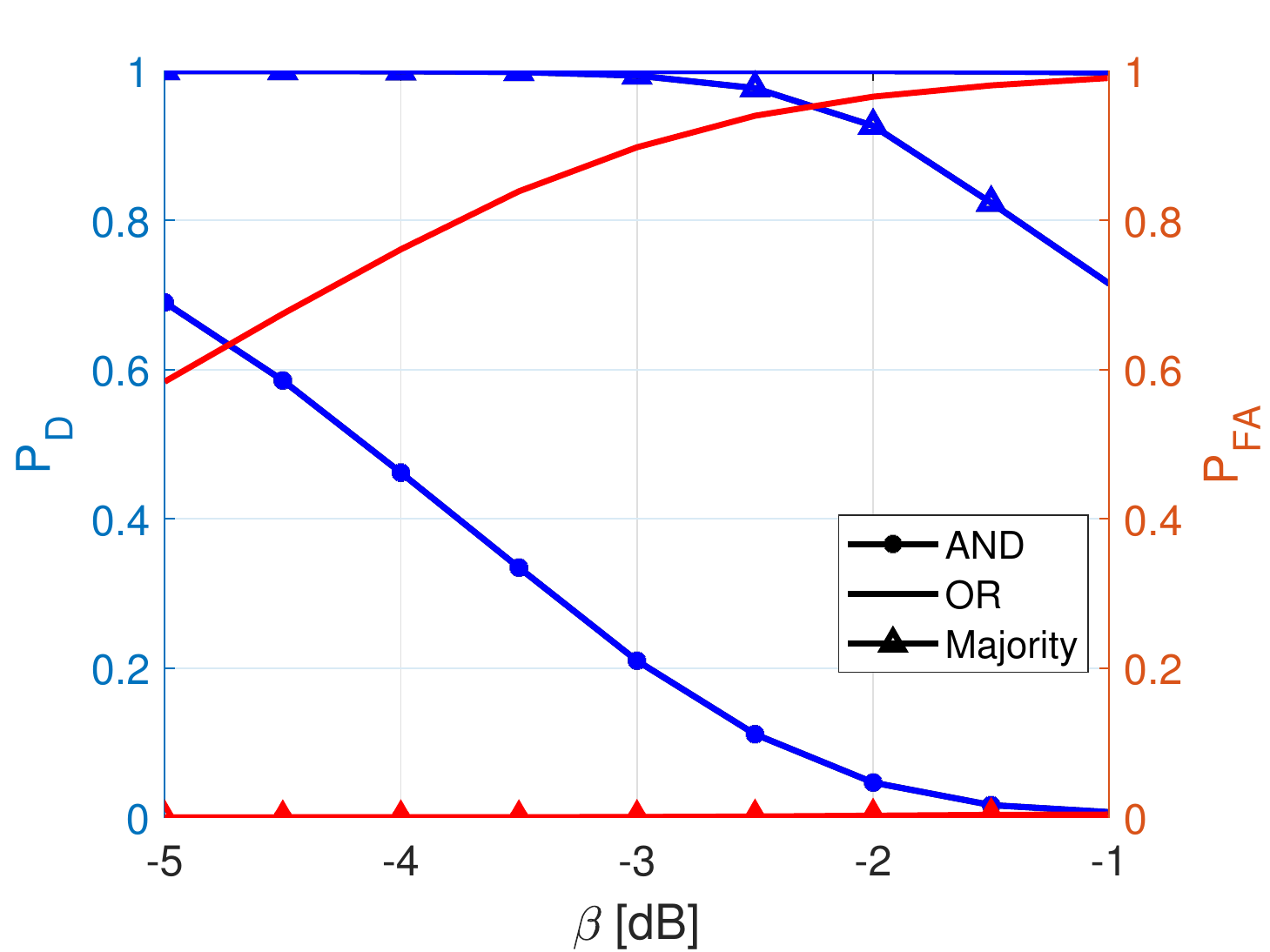} }
		\hfill
		\subfigure[]{
			\label{fig:fusion_3}
			\includegraphics[width=0.3\textwidth]{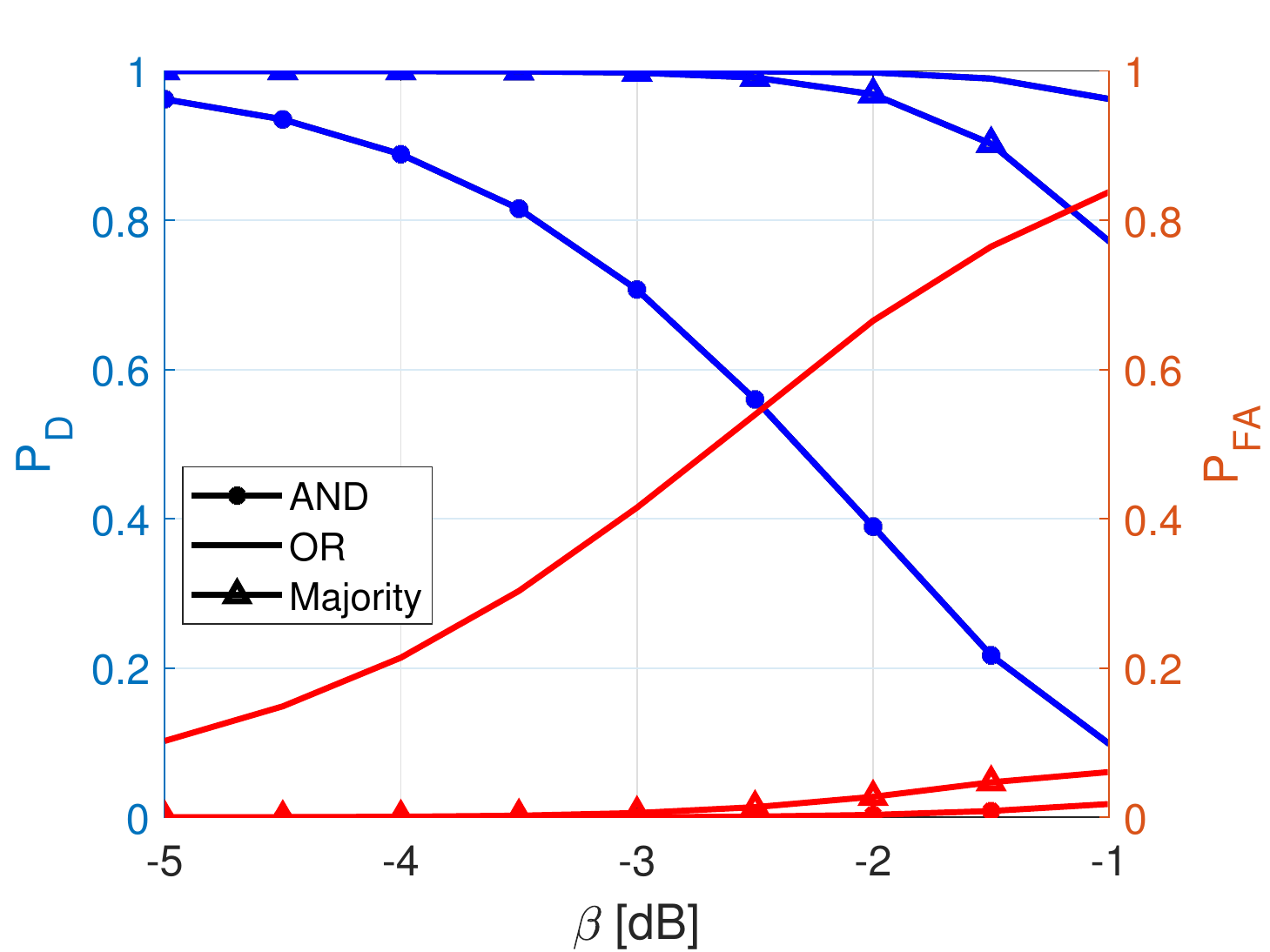} }
	\end{center}
\vspace{-0.5cm}
	\caption{Empirical probability density functions for estimated NPSDSs for three different $N$, $T$ assignments, (a) $N=6$, $T=10$, (b) $N=10$, $T=10$ and (c) $N=6$, $T=20$. The blue colored lines show the joint spoofing detection probability, while the red colored lines show the joint false alarm probability.}
	\label{fig:fusions}
	\vspace{-0.5cm}
\end{figure*}

Numerical analyses can be given in two phases.  We assume that  $\sigma_{w_i}^2=1$, $P_i=1$.  Note that, modulation type and power parameters do not have direct effect on the detection performance, since they only influence the value of NPSDS. The main performance parameters for the numerical analyses are the number of channel uses, $\alpha_i$ and $\beta_{i}$ values along with preferred fusion method.

In the first phase, we analyze the  individual detection probabilities at satellites. We assume  $T=10$.  Fig. \ref{fig:ROCs} illustrates the ROC curves for the detection performance at the $i^{\text{th}}$ satellite. In Fig. \ref{fig:ROC1}, we observe the detection probability and false alarm probability at the $i^{\text{th}}$ satellite for $0<\beta_i\leq1$. The NPSDS is the sampled power spectral density shifted in line with the Doppler shift observed at the satellite. Considering this, Fig. \ref{fig:ROC1} corresponds to the case, where the Doppler shift observed from Alice is higher than the Doppler shift observed from Eve. Therefore, NPSDS values at Eve are higher than the Alice. Since $\beta_{i}=\frac{\Theta_{si}}{\Theta_{ai}}$, as it approaches 1, differentiating Eve and Alice becomes unattainable at the $i^{\text{th}}$ satellite. Fig. \ref{fig:ROC2} shows the spoofing detection probability and false alarm probability values at $i^{\text{th}}$ satellite for $1\leq\beta_i$. Conversely, this case correspond to the higher Doppler frequency shift observations for the incoming signals from Eve. Comparing Fig. \ref{fig:ROC1} to Fig. \ref{fig:ROC2}, higher detection probability of attack can be observed when Eve has a higher NPSDS than Alice.

In the second phase, considering $\beta_{i}=\beta$ values, and $N$, $T$ configurations, each satellite determines their individual optimum detection threshold by Golden search algorithm as previously utilized in \cite{golden}. After giving their individual decisions, a decision vector is fed into the fusion center.  The results related with the fusion decision are presented as in the Fig. \ref{fig:fusions}. Since only single positive detection decision is enough for the positive decision in OR rule, the positive decision becomes easier than the other methods. AND rule requires all decisions to be positive to jointly decide spoofing. Therefore, the spoofing detection probability is the highest in OR rule, followed by the majority rule and lastly lowest in the AND rule. Considering the joint effects of the spoofing detection probability and the false alarm probability, the majority rule  outperforms other fusion mechanisms by allowing a high spoofing detection probability and low false alarm probability. Comparing the results in Fig. \ref{fig:fusion_1}, \ref{fig:fusion_2} and \ref{fig:fusion_3}, we can observe that changing  the number of observer satellites, $N$, have different effects on the fusion mechanisms. For instance, increasing $N$ would decrease the false alarm probability of the AND rule and the majority rule, while this change increases the false alarm probability of the OR rule. Since the estimation quality is strictly related with $T$, as $T$ increases, the spoofing detection probability increases, and the false alarm probability decreases for all fusion rules. 

\section{Conclusion}
In this paper, we have proposed a PLA mechanism specifically designed for the ISLs first time in the literature. We utilize Doppler frequency measurements at different observer satellites as a unique source of identity. By comparing the reference NPSDS values with the measured  NPSDS values, each satellite makes a decision about the identity of the transmitter. The spoofing detection and false alarm probabilities for each decision are analytically obtained. The performance of AND, OR, and majority rules are compared for decision fusion. The numerical analyses show that the majority rule provides the best decision performance among the considered rules. The high detection performance of the majority rule indicates that the applicability of the Doppler-frequency shift based PLA method of ISLs. For future work, we aim to analyze the integration of Doppler based PLA  with existing PLA mechanisms.

\balance

\bibliographystyle{IEEEtran}
\bibliography{references}

\end{document}